\documentclass[a4paper,10pt,twoside]{cpc-hepnp}

\usepackage{multicol}
\usepackage{graphicx}
\usepackage{booktabs}
\usepackage{amssymb,bm,mathrsfs,bbm,amscd}
\usepackage[tbtags]{amsmath}
\usepackage{lastpage}
\usepackage{CJK}

\newcommand{\BR}{{\cal B}}
\newcommand{\ra}{\rightarrow}

\begin{document}

\fancyhead[c]{\small Chinese Physics C~~~Vol. XX, No. X (201X)
XXXXXX} \fancyfoot[C]{\small MMDDYY-\thepage}

\footnotetext[0]{Received xx xx 2015}


\title{Search for $\psi(4S)\to\eta J/\psi$ in
 $B^\pm \ra \eta J/\psi K^\pm$  and
$e^+ e^- \to \eta J/\psi$ processes
\thanks{Supported in part by
the Fundamental Research Funds for the Central
Universities YWF-14-WLXY-013, National Natural Science Foundation of China (11575017)
and CAS center for Excellence in Particle Physics (China)}}

\author{%
\quad Gao Xu-yang$^{1}$%
\quad Wang Xiao-long$^{2,3}$
\quad Shen Cheng-ping$^{1;1)}$\email{shencp@ihep.ac.cn}
}
\maketitle

\address{%
$^1$ School of Physics and Nuclear Energy Engineering, Beihang University\\
Xueyuan Road No.37, Haidian District, Beijing 100191, China\\
$^2$ Institute of High Energy Physics, Chinese Academy of Sciences, Beijing 100049, China\\
$^3$CNP, Virginia Polytechnic Institute and State University, Blacksburg, Virginia 24061\\
}

\begin{abstract}
We search for the $\psi(4S)$ state in the $B^\pm \ra \eta J/\psi K^\pm$ and
 $e^+e^- \ra \eta J/\psi$ processes
based on the Belle measurements with the assumed mass $M=(4230\pm8)$ MeV/$c^2$
and width $\Gamma=(38\pm12)$ MeV.
No significant signal is observed in the $\eta J/\psi$ mass spectra.
The 90\% confidence level  upper limit on the product branching fraction
$\BR(B^\pm \ra \psi(4S)K^\pm)\BR(\psi(4S) \ra \eta J/\psi)<6.8\times 10^{-6}$
 is obtained in the $B^\pm \ra \eta J/\psi K^\pm$ decays.
By assuming the partial width of $\psi(4S) \to e^+e^-$ to be 0.63 keV,
a branching fraction limit $\BR(\psi(4S) \ra \eta J/\psi) < 1.3\%$
is obtained at the 90\% confidence level
in $e^+e^- \ra \eta J/\psi$, which
is consistent with the theoretical prediction.
\end{abstract}



\footnotetext[0]{\hspace*{-3mm}\raisebox{0.3ex}{$\scriptstyle\copyright$}}%

\begin{multicols}{2}

\section{Introduction}

Potential models predict some charmonium states
above the $D\bar{D}$ threshold, but the number of observed states
in experiments is more than the number predicted.
The states which have been seen outwith the theoretical
predictions are normally referred to
as exotic states or XYZ particles.
Many XYZ states have been announced in various processes,
for example, the observation of $X(3872)$ in $B$ decays~\cite{3872},
the $Y(4260)$~\cite{4260}, $Y(4360)$~\cite{4360} and $Y(4660)$~\cite{4660} in
$e^+e^-$ annihilation, and the $X(3915)$~\cite{3915} observed in the two-photon process.

On the other hand, there are still some charmonium states predicted
by the potential models which have not yet been observed experimentally,
especially in the mass region higher than 4 GeV/$c^2$, such as $\eta_c(3S)$, $\eta_c(4S)$, $\psi(4S)$ and $\psi(5S)$.
To some degree, some XYZ states are regarded as candidates for these unfound predicted states.

Searching for these missing predicted states is very helpful to test the potential models.
When checking the mass spectra of the observed charmonia with spin-parity $J^{PC} = 1^{--}$ and
comparing them with those of the corresponding bottomonia, there might be a charmonium state $\psi(4S)$
at about 4.2 GeV/$c^2$ compared to the $\Upsilon(4S)$ state~\cite{Prediction of a missing higher charmonium}.
The authors in Ref.~\cite{Prediction of a missing higher charmonium} predicted that
this missing charmonium state has a mass of 4.263 GeV/$c^2$ and a very narrow width. As a state with
the same spin-parity $1^{--}$, the Y(4220)~\cite{bes3_omegachicj} may be a good candidate for the $\psi(4S)$ state.

Recently, the BESIII Collaboration performed a study on the decay $e^+e^- \ra \omega\chi_{cJ}~ (J = 0,1,2)$~\cite{bes3_omegachicj},
where the Born cross sections at 9 energy points were measured. When using a Breit-Wigner (BW) function to fit the experimental data of $e^+e^- \ra \omega\chi_{c0}$, a resonant structure with mass $M = (4230 \pm 8)$ MeV/$c^2$ and width $\Gamma = (38 \pm 12)$ MeV was observed with a statistical significance more than 9$\sigma$.
However, for the remaining processes $e^+e^- \ra \omega\chi_{c1}$ and $e^+e^- \ra \omega\chi_{c2}$, there were no significant signals.

To understand this novel phenomenon, different explanations of this resonance were given,
which included a tetraquark state~\cite{tetra}, the missing higher charmonium state $\psi(4S)$~\cite{missing higher charmonium calculate},
and the known charmonium resonance $\psi(4160)$~\cite{psi4160}.

The authors in Ref.~\cite{missing higher charmonium calculate}
checked the thresholds of $\omega\chi_{c0}$, $\omega\chi_{c1}$ and $\omega\chi_{c2}$, which are 4.197 GeV/$c^2$,
4.293 GeV/$c^2$ and 4.338 GeV/$c^2$, respectively.
The central mass of $\psi(4S)$ is just above the $\omega\chi_{c0}$ threshold and below the $\omega\chi_{c1,2}$ thresholds.
Accordingly the newly observed structure in $e^+e^- \ra \omega\chi_{c0}$ could be
the missing charmonium $\psi(4S)$ state, and the $e^+e^- \ra \omega\chi_{c1,2}$ processes are kinematically forbidden to $\psi(4S)$.
Stimulated by this, the authors estimated
the meson loop contribution to $\psi(4S)\to \omega \chi_{c0}$ and found the evaluation
can overlap with the experimental data in a reasonable parameter range.
As a typical transition accessible by experiment, the decay $\psi(4S) \ra \eta J/\psi$ similar to $\psi(4S) \ra \omega \chi_{c0}$ can occur.
So the authors in Ref.~\cite{missing higher charmonium calculate} also extended the
theoretical calcution to $\psi(4S) \ra \eta J/\psi$ and predicted the upper limit on
the branching fraction of $\psi(4S) \ra \eta J/\psi$ to be
less than $1.9 \times 10^{-3}$ via the hadronic loop mechanism~\cite{missing higher charmonium calculate}.

As indicated in Ref.~\cite{missing higher charmonium calculate}, the predicted upper limit of
$\psi(4S) \ra \eta J/\psi$ can be accessible at Belle and the forthcoming BelleII. We noticed that
the Belle experiment previously measured the $B^\pm \ra \eta J/\psi K^\pm$~\cite{B_decay}
and $e^+e^- \ra \eta J/\psi$~\cite{isr_decay} processes, where the $\eta J/\psi$ invariant mass distributions
were given. Hence, in this work we fit the $\eta J/\psi$ mass spectra from the
$B^\pm \ra \eta J/\psi K^\pm$ and $e^+e^- \ra \eta J/\psi$ processes
to search for the $\psi(4S)$ state. The experimental measurements
can be taken as a test of the theoretical calculation.

This work is organized as follows. We present the
detailed fit results to the  $\eta J/\psi$ mass spectra from
$B^\pm \ra \eta J/\psi K^\pm$ and $e^+e^- \ra \eta J/\psi$
processes with the $\psi(4S)$ state included in Sec. 2 and Sec. 3.
If no clear $\psi(4S)$ signal is observed, the branching fraction limits
at the 90\% confidence level (C.L.) will be given with the systematic errors included.
The last section ends with the conclusion and discussion.

\section{Search for $\psi(4S)$ in $B$ Decays}

Using $772\times 10^6$ $B\bar{B}$ pairs collected with the Belle detector,
the decays $B^\pm \ra \eta J/\psi K^\pm$ were studied
to search for a new narrow charmonium(-like) state $X$
in the $\eta J/\psi$ mass spectrum, where the
$J/\psi$ and $\eta$ mesons were reconstructed by
a lepton-pair $\ell^+\ell^-$ ($\ell=e,~\mu$) and two photons~\cite{B_decay}.
Except for the known $\psi'\to \eta J/\psi$ decay,
no significant narrow excess was  found in the $\eta J/\psi$
mass spectrum.

Figure~\ref{B_decay_figure} shows the $\eta J/\psi$
mass distribution of interest after all the event selection requirements
are applied.
A binned maximum likelihood fit to the $\eta J/\psi$ mass distribution
is performed to extract the signal and background yields.
A BW function (mass and width fixed at 4.23 GeV/$c^2$ and 38 MeV~\cite{missing higher charmonium calculate})
is convolved with a Gaussian
function (the mass resolution is about 11 MeV/$c^2$) as
the $\psi(4S)$ signal shape and a second polynomial
function is taken as the background shape.
The fit range and results to the $\eta J/\psi$
mass spectrum are shown in Fig.~\ref{B_decay_figure}.

From the fit, we obtain $5.9 \pm 5.5$ signal events, with a
statistical significance of $0.9\sigma$, from the difference
of the logarithmic likelihoods, $-2\ln(\mathcal{L}_0/\mathcal{L}_{\rm max})$, taking
the difference in the number of degrees of freedom ($\Delta
\hbox{ndf}=1$) in the fits into account, where $\mathcal{L}_0$ and $\mathcal{L}_{\rm
max}$ are the likelihoods of the fits without and with a resonance
component, respectively.

We determine a Bayesian 90\% C.L. upper limit
on the number of $\psi(4S)$ signal events ($N_{\rm sig}$)
by finding the value $N^{\rm UP}_{\rm sig}$ such that
$
\int_{0}^{N^{\rm UP}_{\rm sig}} \mathcal{L} dN_{\rm sig}/\int_{0}^{\infty} \mathcal{L} dN_{\rm sig}=0.90,
$
where $N_{\rm sig}$ is the number of $\psi(4S)$ signal events and $\mathcal{L}$ is
the value of the likelihood as a function of $N_{\rm sig}$.
To take into account the systematic uncertainty, the above likelihood
is convolved with a Gaussian function whose width equals the total
systematic uncertainty described below. The upper limit on the number of
$\psi(4S)$ signal events is 22.7 at 90\% C.L.

\begin{center}
\includegraphics[width=8cm]{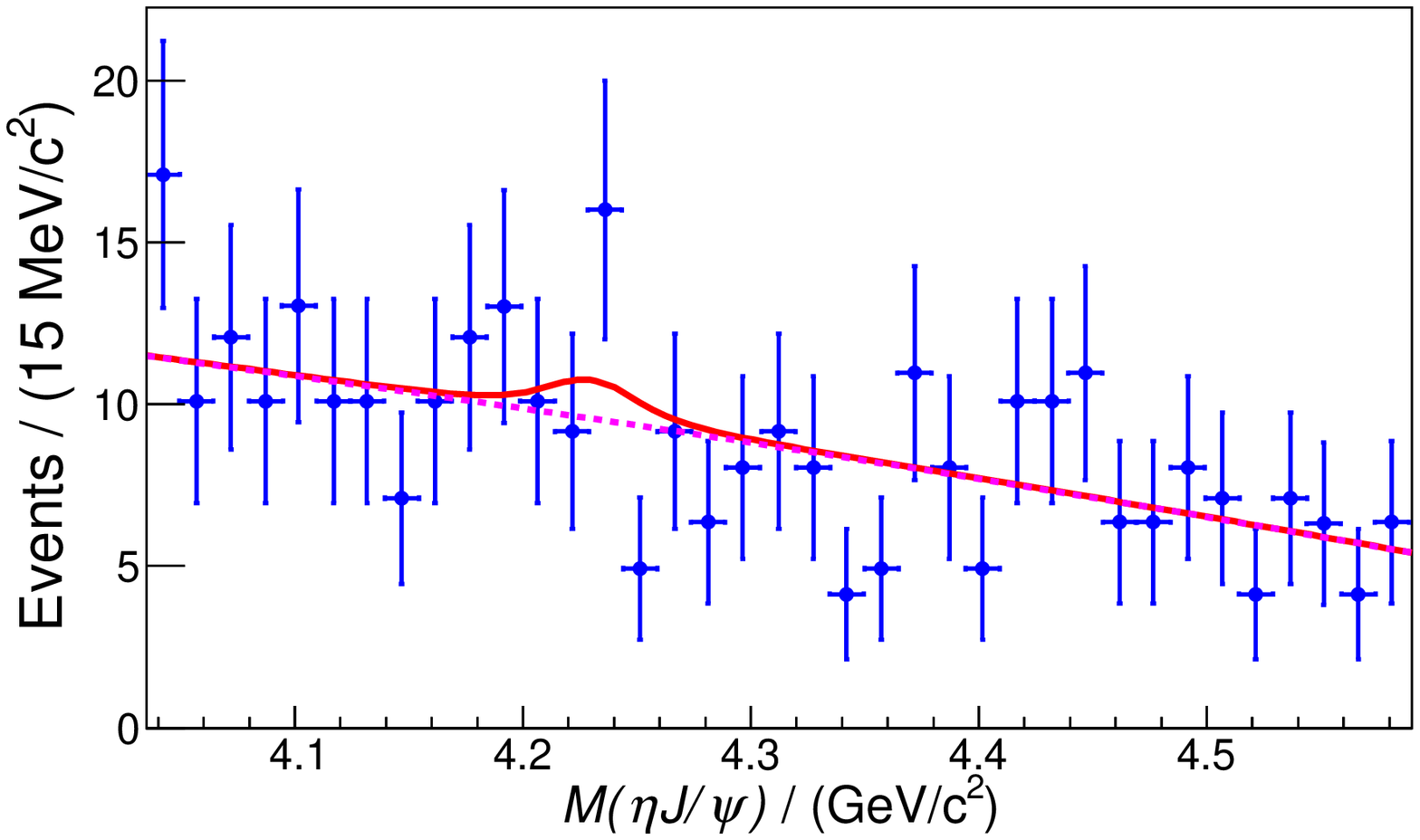}
\figcaption{\label{B_decay_figure} (color online)
The $\eta J/\psi$ invariant mass distribution
from $B^\pm \ra \eta J/\psi K^\pm$ decays.
The dots with error bars are from data, the solid
curve is the best fit for the total signal and the
dotted line shows the fitted background shape.}
\end{center}

There are several sources of systematic error for the
branching fraction measurement. Most of the systematic errors are the same
as those in Ref.~\cite{B_decay}, except that the dominant uncertainty associated with the fitting procedure
is different, which is estimated by changing
the order of the background polynomial, the range
of the fit, the $\psi(4S)$ mass and width by $\pm1\sigma$.
Finally, the uncertainty due to the fitting procedure is 11\%.
Assuming all the sources are independent
and adding them in quadrature, the final total systematic uncertainties
are summarized in Table~\ref{B_decay_uncertainty}.

\begin{center}
\tabcaption{\label{B_decay_uncertainty}
Relative systematic errors (\%) on the
product of the branching fraction
$\BR(B^\pm\ra\psi(4S)K^\pm)\BR(\psi(4S)\ra\eta J/\psi)$.}
\footnotesize
\begin{tabular*}{80mm}[t]{lc}
\hline\hline
     Source   & Relative error ($\%$)    \\
\hline
Tracking efficiency                                                &  1.1        \\
Lepton identification                                              &  2.4         \\
Charged kaon identification                                        &  1.4         \\
$\eta \ra \gamma\gamma$ efficiency                                 &  3.0         \\
Signal MC simulation statistics                                          &  0.5         \\
Secondary $\BR$                                                        &  0.7         \\
$N_{B\bar{B}}$                                                     &  1.4         \\
Fitting procedure  &  11        \\
\hline
Total                                                              &  12        \\
\hline
\bottomrule
\end{tabular*}
\end{center}

The 90\% C.L. upper limit is set on the product branching fraction
$\BR(B^\pm \ra \psi(4S)K^\pm)\BR(\psi(4S) \ra \eta J/\psi)$ using
\begin{eqnarray*}
\label{eq2}
B^{\rm UP}=\frac{N_{\rm sig}^{\rm UP}}{N_{B\bar{B}} \times \epsilon \times \BR(J/\psi \ra \ell^+ \ell^-) \times \BR(\eta \ra \gamma \gamma)},
\end{eqnarray*}
where $N_{\rm sig}^{\rm UP}$, $N_{B\bar{B}}$, $\epsilon$,
$\BR(J/\psi \ra \ell^+ \ell^-)$ and $\BR(\eta \ra \gamma \gamma)$
are the upper limit on the number of
$\psi(4S)$ signal events at 90\% C.L.,
the number of $B\bar{B}$ pairs, the corrected detection efficiency
of 9.23\% at 4.23 GeV/$c^2$ obtained from the fitted efficiency curve using
the efficiencies at $\psi'$, $\psi(4040)$ and $\psi(4160)$ points~\cite{B_decay},
the branching fractions of $J/\psi$ to lepton pair and
$\eta$ to two photons~\cite{PDG},
respectively.  Finally, the 90\% C.L. upper limit on the product branching fraction
$\BR(B^\pm \ra \psi(4S)K^\pm)\BR(\psi(4S) \ra \eta J/\psi)$ is
found to be $6.8\times 10^{-6}$.

\section{Search for $\psi(4S)$ in $e^+e^- \to \eta J/\psi$}

The cross section for $e^+e^- \to \eta J/\psi$ between $\sqrt{s}=3.8$
and 5.3~GeV was measured using 980~fb$^{-1}$ of Belle data,
where the $\eta$ was reconstructed with its $\gamma \gamma$ and $\pi^+ \pi^- \pi^0$
decays and $J/\psi$ was reconstructed via its leptonic decays.
Two distinct resonant structures, the $\psi(4040)$ and $\psi(4160)$, were
observed~\cite{isr_decay}.

To obtain the transition rates of $\psi(4040)$ and $\psi(4160)$ to the
$\eta J/\psi$ final state, an unbinned maximum likelihood fit was performed to the
$\eta J/\psi$ mass spectra from the signal
candidate events and the $\eta$ and $J/\psi$ sideband events
simultaneously~\cite{isr_decay}. The fit to the signal
events includes two coherent $P$-wave BW functions
convolved by the effective
luminosity and efficiency curve for $\psi(4040)$ and $\psi(4160)$ signals
and an incoherent second-order polynomial background; the fit to the
sideband events includes the same background function only.
Due to the low statistics, the masses and widths of the $\psi(4040)$ and $\psi(4160)$
were fixed~\cite{pdg2012} and
the effects of mass resolution were small and therefore were neglected~\cite{isr_decay}.

Similarly here, to obtain the transition rate of $\psi(4S)$ to
$\eta J/\psi$ final state, a binned maximum likelihood fit
with three coherent $P$-wave BW functions for $\psi(4040)$,
$\psi(4160)$ and $\psi(4S)$ is applied to the $e^+e^- \to \eta J/\psi$ cross sections directly,
as shown in Fig.~\ref{isr_result}.
In the fits, besides the masses and widths of $\psi(4040)$ and $\psi(4160)$
are fixed~\cite{pdg2012}, the $\psi(4S)$ parameters are also fixed~\cite{missing higher charmonium calculate}.
Figure~\ref{isr_result_figure} and  Table~\ref{isr_result}
show the fit results. There are four solutions with equally good fit quality.
The results of $\BR\Gamma_{e^+e^-}$ for $\psi(4040)$ and $\psi(4160)$
are consistent with the published results within errors~\cite{isr_decay}.
The significance of the $\psi(4S)$ is estimated by comparing the likelihood of fits
with and without $\psi(4S)$ included. We obtain a statistical
significance of $2.6\sigma$. The most conservative
upper limit with the systematic errors included
on $\BR(\psi(4S)\to\eta J/\psi)\Gamma^{\psi(4S)}_{e^+e^-}$
is obtained to be 8.2 eV at 90\% C.L., which corresponds to Solution III  in Fig.~\ref{isr_result_figure}(c).

Most of the systematic errors
in the $\BR(\psi(4S)\rightarrow \eta J/\psi)\Gamma_{e^+e^-}^{\psi(4S)}$
measurement are the same as those in Ref.~\cite{isr_decay}
except the dominant
systematic error from fit uncertainty (64\%), which includes
the uncertainties on the mass and width of $\psi(4S)$ state
by changing the nominal values by 1$\sigma$~\cite{missing higher charmonium calculate},
and the fit range.
Assuming all the sources are independent and adding them in
quadrature, we obtain total systematic error in $\BR
\Gamma_{e^+ e^-}$ of 65\% for $\psi(4S)$, as shown in Table~\ref{isr_decay_uncertainty}.

\begin{center}
\includegraphics[width=4.4cm]{3r-1.epsi}
\put(-30,50){\bf \large{(a)}}
\includegraphics[width=4.4cm]{3r-2.epsi}
\put(-30,50){\bf \large{(b)}}

\includegraphics[width=4.4cm]{3r-3.epsi}
\put(-30,50){\bf \large{(c)}}
\includegraphics[width=4.4cm]{3r-4.epsi}
\put(-30,50){\bf \large{(d)}}
\vspace{2 mm}
\figcaption{\label{isr_result_figure} (color online)
The $e^+ e^- \to \eta J/\psi$ cross section distributions and the fit results described in the text.
There are four solutions with the coherent $\psi(4040)$, $\psi(4160)$
and $\psi(4S)$ signals.
The curves show the best fit for the measured cross sections and the contribution from each
BW component. The interference term for each solution
 is not shown. }
\end{center}

\begin{table*}[!htb]
\caption{\label{isr_result}
Results of the fits to the $e^+e^- \to \eta J/\psi$ cross sections
using three coherent resonances: $\psi(4040)$, $\psi(4160)$
and $\psi(4S)$. The errors are statistical only.
$M$, $\Gamma$, and $\BR\Gamma_{e^+e^-}$ are the
mass (in MeV/$c^2$), total width (in MeV), product of the
branching fraction to $\eta J/\psi$ and the $e^+e^-$ partial width (in
eV), respectively. $\phi_1$ is the relative phase between the
$\psi(4040)$ and $\psi(4160)$ (in degrees) and $\phi_2$ is the relative phase between the
 $\psi(4160)$ and $\psi(4S)$ (in degrees).
}
\begin{center}
\centering
\tabcolsep=5pt
\begin{tabular}[t]{c|cccc}
\hline\hline
     & Solution I   & Solution II   & Solution III  & Solution IV   \\
\hline
$M_{\psi(4040)} $    &  \multicolumn{4}{c}{4039(fixed)}         \\
$\Gamma_{\psi(4040)} $     & \multicolumn{4}{c}{80(fixed)}   \\
$\BR(\psi(4040)\rightarrow \eta J/\psi)\Gamma_{e^+e^-}^{\psi(4040)}$    & 6.1 $\pm$ 1.1    & 11.8 $\pm$ 1.4   & 12.2 $\pm$ 1.5  & 6.3 $\pm$ 1.1  \\
$M_{\psi(4160)} $  &   \multicolumn{4}{c}{4153(fixed)}          \\
$\Gamma_{\psi(4160)} $    &   \multicolumn{4}{c}{103(fixed)}        \\
$\BR(\psi(4160)\rightarrow \eta J/\psi)\Gamma_{e^+e^-}^{\psi(4160)}$    & 6.3 $\pm$ 1.6  & 16.7 $\pm$ 2.2  & 20.4 $\pm$ 2.4 & 7.7 $\pm$ 1.7  \\
$M_{\psi(4S)} $      &   \multicolumn{4}{c}{4230(fixed)}          \\
$\Gamma_{\psi(4S)} $     &   \multicolumn{4}{c}{38(fixed)}         \\
$\BR(\psi(4S)\rightarrow \eta J/\psi)\Gamma_{e^+e^-}^{\psi(4S)}$      & 0.8 $\pm$ 0.6  & 0.9 $\pm$ 0.7  & 4.5 $\pm$ 1.3 & 4.0 $\pm$ 1.2  \\
$\phi_1$       & 320 $\pm$ 12 & 258 $\pm$ 6 & 262 $\pm$ 5   & 324 $\pm$ 12    \\
$\phi_2$       & 171 $\pm$ 16 & 117 $\pm$ 17 & 142 $\pm$ 8   & 197 $\pm$ 12    \\
\hline\hline
\end{tabular}
\end{center}
\end{table*}

\begin{center}
\tabcaption{ \label{isr_decay_uncertainty}
Relative systematic errors (in \%) in the
$\BR(\psi(4S)\rightarrow \eta J/\psi)\Gamma_{e^+e^-}^{\psi(4S)}$
measurement.}
\footnotesize
\begin{tabular*}{80mm}[t]{lc}
\hline\hline
     Source   & Relative error ($\%$)    \\
\hline
Particle identification                                                         &  5.5          \\
Tracking efficiency                                                 &  1.4          \\
Photon reconstruction                                               &  4.0          \\
$J/\psi$, $\eta$ mass etc requirements                                      &  2.6          \\
Luminosity measurement                                              &  1.4          \\
MC generator                                                        &  1.0          \\
Trigger simulation                                                  &  2.0          \\
Intermediate decay branching fractions                              &  1.6          \\
Signal MC simulation statistics                                     &  0.2          \\
Fit uncertainty                                                     &  64         \\
\hline
Total                                                               &  65         \\
\hline\hline
\end{tabular*}
\end{center}

In Refs.~\cite{dong, Gamma}, the partial width of $\psi(4S) \to e^+ e^-$ was estimated,
i.e., $\Gamma(\psi(4S) \ra e^+e^-) = 0.63$ keV~\cite{dong} and $\Gamma(\psi(4S) \ra e^+e^-) = 0.66$ keV~\cite{Gamma}.
If we take the lower theoretical calculation $\Gamma(\psi(4S) \ra e^+e^-) = 0.63$ keV~\cite{dong},
we can obtain the conservative upper limit
on branching fraction $\BR(\psi(4S)\to \eta J/\psi)$ at 90$\%$ C.L., i.e.,
\centerline{$\BR(\psi(4S) \ra \eta J/\psi) < 1.3\%$,}
which does not contradict the theoretical prediction of $1.9 \times 10^{-3}$~\cite{missing higher charmonium calculate}.

\section{Summary}

In summary,
we search for the $\psi(4S)$ state in the $B^\pm \ra \eta J/\psi K^\pm$ and
$e^+e^- \ra \eta J/\psi$ processes based on the Belle measurements with the assumed mass $M=(4230\pm8)$ MeV/$c^2$
and width $\Gamma=(38\pm12)$ MeV.
The $\eta J/\psi$ mass spectrum from $B$ decays and
the cross sections of $e^+e^- \to \eta J/\psi$ are fitted with the $\psi(4S)$ resonance included
for the first time. No significant signal is observed, and 90\% C.L. upper limits of
$\BR(B^\pm \ra \psi(4S))\BR(\psi(4S) \ra \eta J/\psi K^\pm) < 6.8 \times 10^{-6}$
and $\BR(\psi(4S)\rightarrow \eta J/\psi)\Gamma_{e^+e^-}^{\psi(4S)}<8.2$ eV are obtained.
With the $\Gamma(\psi(4S) \ra e^+e^-) = 0.63$ keV~\cite{dong} as input, we have
$\BR(\psi(4S) \ra \eta J/\psi) < 1.3\%$, which is almost one order higher
in magnitude than the theoretical prediction~\cite{missing higher charmonium calculate}.

The expected integrated luminosity at the BelleII experiment is 50~ab$^{-1}$
in 2024, which is about 50 times the current total integrated luminosity
at Belle.  With this huge data sample, the expected upper limit
on $\BR(\psi(4S) \ra \eta J/\psi)$ will be $1.9 \times 10^{-3}$ if it scales
as 1/$\sqrt{L}$, where $L$ is the integrated luminosity,
and can therefore reach the theoretical prediction level.

\end{multicols}

\vspace{-1mm}
\centerline{\rule{80mm}{0.1pt}}
\vspace{2mm}

\begin{multicols}{2}

\end{multicols}

\clearpage

\end{document}